\documentclass{amsart} 
\usepackage[utf8]{inputenc} 
\usepackage[T1]{fontenc}    
\usepackage{lmodern}        
\usepackage{hyperref}       
\usepackage{url}            
\usepackage{booktabs}       
\usepackage{amsfonts}       
\usepackage{nicefrac}       
\usepackage{microtype}      

\usepackage[linesnumbered,ruled,vlined]{algorithm2e}
\usepackage{amsthm}
\usepackage{amsmath}
\usepackage[conf={restate,no link to proof}]{proof-at-the-end}

\newtheorem{theorem}{Theorem}
\newtheorem{definition}[theorem]{Definition}

\newcommand{\ind}[1]{\ensuremath{\mathbf{1}_{\{#1\}}}}
\DeclareMathOperator*{\argmax}{argmax}
\DeclareMathOperator{\expo}{Expo}

\DeclareMathOperator{\bern}{Bernoulli}

\newcommand{\ex}[2]{{\ifx&#1& \mathbb{E} \else
\underset{#1}{\mathbb{E}} \fi \left[#2\right]}}
\newcommand{\pr}[2]{{\ifx&#1& \mathbb{P} \else
\underset{#1}{\mathbb{P}} \fi \left[#2\right]}}
\newcommand{\var}[2]{{\ifx&#1& \mathsf{Var} \else
\underset{#1}{\mathsf{Var}} \fi \left[#2\right]}}

\keywords{Permute-and-Flip, Exponential Mechanism, Report Noisy Max}

\title[Permute-and-flip is Noisy Max]{The Permute-and-Flip Mechanism is Identical to Report-Noisy-Max with Exponential Noise}

\author[Z.~Ding]{Zeyu Ding}	
\address{Penn State University, State College, PA 16801}	

\author[D.~Kifer]{Daniel Kifer}	
\address{Penn State University, State College, PA 16801}	

\author[S.~Saghaian N. E.]{Sayed M. Saghaian N. E.}	
\address{Penn State University, State College, PA 16801}	

\author[T.~Steinke]{Thomas Steinke}	
\address{Google, Mountain View, CA 94043}	

\author[Y.~Wang]{Yuxin Wang}	
\address{Penn State University, State College, PA 16801}	

\author[Y.~Xiao]{Yingtai Xiao}	
\address{Penn State University, State College, PA 16801}	

\author[D.~Zhang]{Danfeng Zhang}	
\address{Penn State University, State College, PA 16801}	

\begin{document}

\begin{abstract}
The permute-and-flip mechanism is a recently proposed differentially private selection algorithm that was shown to outperform the exponential mechanism. In this paper, we show that permute-and-flip is  equivalent to the well-known report noisy max algorithm with exponential noise. 
\end{abstract}

\maketitle

\section{Introduction}\label{sec:intro}
 Differentially private selection is an important mechanism in the differential privacy toolbox. For example, it is a key component in many privacy preserving algorithms for synthetic data generation \cite{mwem}, ordered statistics \cite{BlockiDB16}, quantiles \cite{smithconverge}, frequent itemset mining \cite{BhaskarDFP}, hyperparameter tuning \cite{LiuPSPC} for statistical models, etc.
 
At its core, private selection is a way of making a choice among (usually discrete) outcomes based on a confidential dataset $D$. Specifically, given a set of possible outcomes $\omega_1,\omega_2,\dots,\omega_k$, one designs a quality function $q$ so that $q(D,\omega_i)$ measures the utility of choosing $\omega_i$ when the true dataset is $D$. The goal is to identify the $\omega_i$ that likely maximizes $q(D,\omega)$. For example, if $\omega_1,\dots, \omega_k$ are products to promote and $D$ is a confidential database, $q(D,\omega_i)$ could measure expected profit of promoting item $\omega_i$. Or, if $\omega_1,\dots, \omega_k$ are non-disjoint subsets of possible features for a machine learning model, $q(D,\omega_i)$ could measure anticipated performance of the model if only features from $\omega_i$ are used for training.

A very popular algorithm for performing differentially private selection is the Exponential Mechanism \cite{exponentialMechanism}, which is a special case of another popular selection algorithm called Report Noisy Max \cite{diffpbook}. There are three well-established versions of Report Noisy Max: one which uses noise from a Gumbel distribution (and which, in fact, is identical to the exponential mechanism \cite{diffpbook}), another one that uses Laplace noise \cite{diffpbook}, and a third one that uses Exponential noise \cite{diffpbook}.\footnote{Some sources erroneously claimed that Report Noisy Max with Exponential noise is equivalent to the exponential mechanism (these claims have since been corrected.)} Recently, a new selection algorithm called Permute-and-Flip was proposed by McKenna and Sheldon \cite{paf} with the exciting guarantee that its expected utility dominates the exponential mechanism. In this paper, we present a full proof that the Permute-and-Flip mechanism is identical to Report Noisy Max with Exponential noise.

In Section \ref{sec:prelim}, we present background on differential privacy and private selection mechanisms. In Section \ref{sec:equivalence} we prove the main result of the paper
, and we discuss related work in Section \ref{sec:related}. We present conclusions in Section \ref{sec:conc}.

\section{Background}\label{sec:prelim}
Differential privacy is by now considered a de facto standard for protecting confidentiality of user data in statistical applications and has been adopted by organizations such as Google \cite{rappor,prochlo}, Apple \cite{applediffp}, Microsoft  \cite{DingKY17} and the U.S. Census Bureau \cite{ashwin08:map,Haney:2017:UCF,abowd18kdd}. Formally, it is defined as follows:

\begin{definition}[Pure Differential Privacy \cite{dwork06Calibrating}]
Given a privacy loss budget $\varepsilon>0$, a randomized algorithm $M$  satisfies (pure) $\varepsilon$-differential privacy if, for each possible output $\omega$ and all pairs of datasets $D_1, D_2$ that differ on the addition or removal of the record(s) of one person, the following holds:
\begin{equation}\label{eq:dpdef}
    \pr{}{M(D_1)=\omega} \leq e^\varepsilon \cdot \pr{}{M(D_2)=\omega}
\end{equation}
where the probability is only over the randomness of $M$.\footnote{This definition applies to algorithms with a discrete output space. More generally, we must quantify over all (measurable) sets of outputs $S$, rather than individual outputs $\omega$, and require $\pr{}{M(D_1) \in S} \le e^\varepsilon \cdot \pr{}{M(D_2) \in S}$.}
\end{definition}
Intuitively, differential privacy guarantees that the output distribution of $M$ does not depend too much on any individual's data. Thus a potential adversary cannot infer much about any individual from the output.

\emph{Private selection} is a specific task, defined as follows. Let $D$ be a private dataset. We are given a (public) set $\Omega=\{\omega_1,\dots,\omega_k\}$ of possible outcomes and a quality function $q$; the value $q(D,\omega_i)$ is interpreted as the utility of selecting $\omega_i$ when the true dataset is $D$. 
The task of private selection is to design a randomized algorithm $M$ that returns some $\omega_i$ that approximately maximizes $q(D,\omega_i)$ while satisfying differential privacy. Specifically, $M$ should optimize the probability that it returns an optimal or near-optimal $\omega_i$ subject to privacy constraints.

The \emph{sensitivity} $\Delta$ of the quality function $q$ is defined as
\begin{equation}
\Delta := \sup_{D_1,D_2} \max_{\omega_i} |q(D_1,\omega_i) - q(D_2,\omega_i)|,
\end{equation}
where the supremum is taken over all pairs of datasets $D_1,D_2$ that differ in the addition/removal of one person's data.

One of the most popular differentially private selection algorithms is called \emph{Report Noisy Max}. Independent noise is added to each value $q(D,\omega_1), q(D,\omega_2), \cdots, q(D,\omega_k)$ and the $\omega_i$ corresponding to the largest noisy value is returned. Three variants -- each with a different noise distribution -- have been studied \cite{diffpbook}:
\begin{enumerate}
\item The first variant, called \textbf{Report Noisy Max with Exponential Noise}, is shown in Algorithm \ref{alg:nm}. It uses the exponential distribution, denoted as Expo$(\lambda)$, which has density function\footnote{The $\lambda$ parameter is known as the rate and its inverse (known as the scale) is proportional to the  standard deviation.} $f(x;\lambda)=\lambda e^{-\lambda x}\ind{x\geq 0}$ and generates nonnegative noise. The algorithm evaluates the quality $q(D,\omega_i)$ of each item, adds independent Expo$(\frac{\varepsilon}{2\Delta})$ noise to each quality score, finds the  $\omega_i$ that has the largest noisy quality score and returns its index $i$.
\item The second variant uses the Laplace distribution Lap$(\alpha)$ with density function\footnote{The $\alpha$ parameter of the Laplace distribution is the scale and is proportional to the standard deviation.} $f(x;\alpha)=\frac{1}{2\alpha} e^{-|x|/\alpha}$. The only difference with the first variant is that instead of adding Expo$(\frac{\varepsilon}{2\Delta})$ noise, it adds Lap$(2\Delta/\varepsilon)$ noise.
\item The third variant uses the Gumbel distribution Gumbel$(\alpha)$ with density function\footnote{The $\alpha$ parameter of the Gumbel distribution is the scale and is proportional to the standard deviation.} $f(x;\alpha)=\frac{1}{\alpha}\exp(-\frac{x}{\alpha} - e^{-x/\alpha})$. Instead of adding Expo$(\frac{\varepsilon}{2\Delta})$ noise, it adds Gumbel$(2\Delta/\varepsilon)$ noise. This variant is known as the \textbf{Exponential Mechanism} \cite{exponentialMechanism} and it is the only variant whose output distribution has a simple formula:
\begin{equation}
\pr{}{M(D) = \omega_i} = \frac{\exp\left(\frac{\varepsilon}{2\Delta}q(D,\omega_i)\right)}{\sum_{j=1}^k \exp\left(\frac{\varepsilon}{2\Delta}q(D,\omega_j)\right)}
\end{equation}
\end{enumerate}

\begin{algorithm}[h!]
\DontPrintSemicolon
\KwIn{ Private dataset $D$, quality function $q$ with sensitivity $\Delta>0$ and set of outcomes $\Omega = \{ \omega_1, \cdots, \omega_k \}$, privacy parameter $\varepsilon>0$}
\For{$i=1,\dots, k$}{
  $v_i \gets q(D, \omega_i) + \expo(\frac{\varepsilon}{2\Delta})$\;
}
\Return $\argmax_i v_i$
\caption{Report Noisy Max with Exponential Noise \cite{diffpbook}}\label{alg:nm}
\end{algorithm}

Recently, a new mechanism, called Permute-and-Flip was proposed \cite{paf} and is shown in Algorithm \ref{alg:paf}. It was proved that the expected error of Permute-and-Flip is never worse than that of the exponential mechanism. First it computes the true best score $q_*=\max_i q(D,\omega_i)$. Each $\omega_i$ is then assigned  a coin that lands heads with probability $\exp\left(\frac{\varepsilon}{2\Delta}(q(D,\omega_i)-q_*)\right)$. The algorithm  sequentially processes the $\omega_i$ in a random order and flips their associated coins. It stops as soon as a coin lands heads and returns the $\omega_i$ corresponding to that coin.

\begin{algorithm}[!ht]
\DontPrintSemicolon
\KwIn{ Private dataset $D$, quality function $q$ with sensitivity $\Delta>0$ and set of outcomes $\Omega = \{ \omega_1, \cdots, \omega_k \}$, privacy parameter $\varepsilon>0$}
$q_* \gets \max_i q(D,\omega_i)$\;
$\pi \gets $ random permutation of $1\dots, k$.\;
\For{$i=1,\dots, k$}{
  $r \gets \pi(i)$\;
  $p \gets \exp\left(\frac{\varepsilon}{2\Delta}(q(D,\omega_r)-q_*)\right)$\;
  \If{$\bern(p)$}{
     \Return $r$\;
  }
}
\caption{Permute-and-flip \cite{paf}}\label{alg:paf}
\end{algorithm}

In this paper, we show that Algorithms \ref{alg:nm} and \ref{alg:paf} are equivalent. That is, even though they are implemented differently, their output distributions are exactly the same.

\section{The Equivalence}\label{sec:equivalence}
We prove the equivalence of the output distributions of Algorithms \ref{alg:paf} and \ref{alg:nm} through a series of intermediate algorithms.
In this section, we denote the true maximum value of the quality function over $D$ as $q_*=\max_i q(D,\omega_i)$.

The first intermediate mechanism is shown in Algorithm \ref{alg:a}. Instead of randomly permuting the elements and flipping a coin, it adds exponential noise to each quality score to get a noisy score $v_i$. It creates a set $S$ where it stores all indexes $i$ for which the noisy quality score $v_i$ exceeds $q_*$. Among all of these, it returns one at random. Note that the set $S$ in Algorithm \ref{alg:a} is never empty -- if $i^*$ is the index of the true max (i.e., $q(D, \omega_{i^*})=q_*$) then the noisy score $v_{i^*}$ is at least $ q_*$ since exponential noise is never negative (and hence $i^*\in S$).

\begin{algorithm}[!ht]
\DontPrintSemicolon
\KwIn{ Private dataset $D$, quality function $q$ with sensitivity $\Delta>0$ and set of outcomes $\Omega = \{ \omega_1, \cdots, \omega_k \}$, privacy parameter $\varepsilon>0$}
$q_* \gets \max_i q(D,\omega_i)$\;
\For{$i=1,\dots, k$}{
  $v_i \gets q(D,\omega_i) + \expo(\frac{\varepsilon}{2\Delta})$\;
}
$S\gets \{i~:~ v_i \geq q_*\}$\;
\Return one random element from $S$ (chosen uniformly at random).
\caption{Intermediate Algorithm $A$}\label{alg:a}
\end{algorithm}

\begin{theoremEnd}[category=equivalence,proof here]{lemma}\label{thm:part1} The output distributions of Algorithms \ref{alg:paf} and \ref{alg:a} are the same.
\end{theoremEnd}
\begin{proofEnd}
Let $X$ be an $\expo(\varepsilon/(2\Delta))$ random variable. We have $\pr{}{X \ge x} = \exp(-\frac{\varepsilon x}{2\Delta})$ for all $x \ge 0$. Using the notation in Algorithm \ref{alg:a}, we note that $\pr{}{i\in S}=\pr{}{v_i \geq q_*}$ is equal to $\pr{}{X \geq q_*-q(D,\omega_i)}=\exp\left(-\frac{\varepsilon}{2\Delta}(q^*-q(D,\omega_i))\right)=\exp\left(\frac{\varepsilon}{2\Delta}(q(D,\omega_i)-q_*)\right)$. Thus, Algorithm \ref{alg:a} is equivalent to an algorithm that, for each $i=1,\dots, k$, adds $i$ to $S$ independently with probability $\exp\left(\frac{\varepsilon}{2\Delta}(q(D,\omega_i)-q_*)\right)$  and then returns a random element of $S$. Now, returning a random element of $S$ is the same as permuting $S$ and returning the first element in the permuted order. Thus, Algorithm $\ref{alg:a}$ is the same as permuting $1,\dots, k$ and returning the first $i$ for which a Bernoulli$\left(\exp\left(\frac{\varepsilon}{2\Delta}(q(D,\omega_i)-q_*)\right)\right)$ random variable equals 1, which is exactly what Permute-and-Flip (Algorithm \ref{alg:paf}) does.
\end{proofEnd}

We make the following modifications to Algorithm \ref{alg:a} to obtain our next intermediate mechanism, shown in Algorithm \ref{alg:b}. First, we cap all the noisy quality scores $v_i$ at $q_*$ (note that in Algorithm \ref{alg:a} it didn't matter by how much $v_i$ exceeded $q_*$). The capped noisy scores are denoted as $v_i^\top$. Then, for all the items whose noisy capped scores are equal to $q_*$ (call this set $S^\prime$), we add exponential noise to the capped values and return the index $i$ of the $\omega_i$ with the largest of these noisy values.  Clearly this is no different than selecting one element of $S^\prime$ uniformly at random, but it allows us to use the memorylessness property of the exponential noise to connect Algorithm \ref{alg:b} with Algorithm \ref{alg:nm}. 

\begin{algorithm}[!ht]
\DontPrintSemicolon
\KwIn{ Private dataset $D$, quality function $q$ with sensitivity $\Delta>0$ and set of outcomes $\Omega = \{ \omega_1, \cdots, \omega_k \}$, privacy parameter $\varepsilon>0$}
$q_* \gets \max_i q(D,\omega_i)$\;
\For{$i=1,\dots, k$}{
  $v^\top_i \gets \min\left\{q_*,~q(D,\omega_i) + \expo(\frac{\varepsilon}{2\Delta})\right\}$\;\label{line:b:qstar}
  $z_i \gets \expo(\frac{\varepsilon}{2\Delta})$
}
$S^\prime\gets \{i~:~ v^\top_i = q_*\}$\;
\Return $\argmax_{i\in S^\prime} ~v^\top_i + z_i$ \tcp*{Same as $\argmax_{i\in S^\prime} ~q_* + z_i$}\label{line:b:return}
\caption{Intermediate Algorithm $B$}\label{alg:b}
\end{algorithm}


Next we show that Algorithms \ref{alg:a} and \ref{alg:b} are equivalent, which implies that Permute-and-Flip (Algorithm \ref{alg:paf}) is also equivalent to Algorithm \ref{alg:b}.
\begin{theoremEnd}[category=equivalence,proof here]{lemma}\label{thm:part2} The output distributions of Algorithms \ref{alg:a} and \ref{alg:b} are the same.
\end{theoremEnd}
\begin{proofEnd}
We first note that the set $S$ in Algorithm \ref{alg:a} and $S^\prime$ in Algorithm \ref{alg:b} have the same distribution, since Algorithm \ref{alg:b} truncates $v_i$ at $q_*$ to get $v_i^\top$. Thus the only difference between the algorithms is how an element from $S$ and $S^\prime$ is chosen. In Algorithm \ref{alg:a}, this happens uniformly at random. In Algorithm \ref{alg:b}, note that $v_i^\top=q^*$ for all $i\in S^\prime$ by construction (Line \ref{line:b:qstar}), so that Line \ref{line:b:return} satisfies $\argmax_{i\in S^\prime} ~v^\top_i + z_i=\argmax_{i\in S^\prime} ~q_* + z_i=\argmax_{i\in S^\prime} z_i$. Since for all $i\in S^\prime$, the $z_i$ are i.i.d.~exponential random variables and ties have probability zero with continuous noise, by symmetry, Line \ref{line:b:return} in Algorithm \ref{alg:b} also returns an element of $S^\prime$ uniformly at random. Hence, these two algorithms' output distributions are identical.
\end{proofEnd}

Finally, we show that Algorithm \ref{alg:b} is equivalent to Report Noisy Max with exponential noise (Algorithm \ref{alg:nm}), which completes the proof of the equivalence.
\begin{theoremEnd}[category=equivalence,proof here]{lemma}\label{thm:part3} The output distributions of Algorithm \ref{alg:b} and Report Noisy Max with exponential noise (Algorithm \ref{alg:nm}) are the same.
\end{theoremEnd}
\begin{proofEnd}
For Algorithm \ref{alg:b}, we define some intermediate variables as follows:
\begin{align*}
v^\dagger_i &= 
\begin{cases}
v_i^\top & \text{ if } v_i^\top < q_*\\
q_* + z_i & \text{ if } v_i^\top = q_*
\end{cases}
\end{align*}
Note the $v_i^\dagger$ variables are independent of each other. We next claim that, for each $i$, the distribution of $v_i^\dagger$ is the same as the distribution of $q(D,\omega_i)+\expo\left(\frac{\varepsilon}{2\Delta}\right)$. This follows from the memoryless property of the exponential distribution: For any $t \in [q(D,\omega_i), ~~ q_*)$, we have
\begin{align*}
\pr{}{v_i^\dagger\leq t } &= \pr{}{v_i^\top \leq t} = 1-\exp\left(-\frac{\varepsilon}{2\Delta}(t-q(D,\omega_i))\right).\\
\intertext{For any $t \geq q_*$, we have}
\pr{}{v_i^\dagger\leq t } &= \pr{}{v_i^\top < q_*} + \pr{}{v_i^\top = q_* ~~\wedge~~ z_i +q_*\leq t}\\
                      &= \left(1-\exp\left(-\frac{\varepsilon}{2\Delta}(q_*-q(D,\omega_i))\right)\right) \\&~~~~~~~~~~~~+ \exp\left(-\frac{\varepsilon}{2\Delta}(q_*-q(D,\omega_i))\right)\left(1-\exp\left(-\frac{\varepsilon}{2\Delta}(t-q_*)\right)\right)\\
                      &= 1-\exp\left(-\frac{\varepsilon}{2\Delta}(t-q(D,\omega_i))\right).
\end{align*}
This proves the claim, as this is precisely the cumulative distribution function of $q(D,\omega_i) + \expo(\frac{\varepsilon}{2\Delta})$. 

Now, noting that $v_i^\dagger \geq q_*> v_j^\dagger$ whenever $i\in S^\prime$ and $j\not\in S^\prime$, we have
\begin{align*}
\argmax_{i\in S^\prime} ~v_i^\top + z_i&=\argmax_{i\in S^\prime} ~q_* + z_i
=\argmax_{i\in S^\prime} ~v^\dagger_i\\
&=\argmax_{i\in\{1,\dots, k\}} ~v^\dagger_i \overset{d}{=} \argmax_{i\in \{1,\dots, k\}} \left(q(D,\omega_i) +\expo(\frac{\varepsilon}{2\Delta})\right)
\end{align*}
where $\overset{d}{=}$ means equality in distribution.
Since the right hand side is Report Noisy Max with exponential noise  (Algorithm \ref{alg:nm}), the proof is done.
\end{proofEnd}

Thus, putting together Lemmas \ref{thm:part1}, \ref{thm:part2}, and \ref{thm:part3}, we have the following result:
\begin{theorem} The permute-and-flip and report-noisy-max-with-exponential-noise algorithms (Algorithms \ref{alg:paf} and \ref{alg:nm} respectively) have the same output distributions.
\end{theorem}


\section{Related Work}\label{sec:related}
The optimality results of McKenna and Sheldon \cite{paf} show that Permute-and-Flip and, thus by our equivalence, also Report Noisy Max with Exponential noise is perferrable over the exponential mechanism for most applications. There are several noteworthy improvements to the noisy max family of algorithms.


Zhang et al. \cite{privgene} found a tighter bound on sensitivity of certain quality functions when used with the exponential mechanism. Later, Dong et al. \cite{dongexpo} found a more general bound: 
$$\Delta=\sup_{\text{neighbors } D_1, D_2} \left(\max_{\omega} (q(D_1, \omega) - q(D_2,\omega))\right) - \left(\min_{\omega^*}(q(D_1, \omega^*) - q(D_2,\omega^*))\right)$$
that can be used with the exponential mechanism. This bound can be plugged directly into the privacy proof of the noisy max algorithm \cite{diffpbook}.


Ding et al.~\cite{freegap} generalized noisy max in a different direction and showed that one can release additional information at no extra privacy cost. Specifically, given the vector of noisy quality scores, in addition to releasing the argmax of this vector, one can also release the ``gap'' -- the difference between the noisy quality scores of the winning index and the second best index. This does not affect the privacy cost. Note that releasing the max of the noisy vector incurs additional privacy cost \cite{statdp}, but the gap does not (if one is interested in an estimate of the maximum quality score, one would just compute the maximum and add Lap$(\Delta/\varepsilon)$ noise to it).

Chaudhuri et al.~\cite{chaudhuri2014large} study a ``large margin'' variant of private selection (also known as GAP-MAX \cite{bun2018composable,bun2019private}). Here we add the assumption that very few outcomes have high quality -- i.e., we assume $|\{\omega_i : q(D,\omega_i) \ge \max_j q(D,\omega_j) - t\}|$ is small for some $t>0$. Under this assumption, it is possible to obtain utility bounds that do not depend on the total number of outcomes $k$.

Liu and Talwar \cite{liu2019private} a variant of the selection problem where the outcomes $\omega_1, \cdots, \omega_k$ are not fixed in advance and are instead the outputs of some differentially private algorithm. 

An online variant of private selection has also been studied. Here the outcomes $\omega_1, \cdots, \omega_k$ arrive in a streaming fashion and the algorithm must decide whether to select each outcome before the next outcome is presented. Thus, rather than selecting the best outcome, the goal is to select some outcome that is above a pre-defined threshold. The Sparse Vector algorithm \cite{diffpbook} provides a solution to this problem. 

Ilvento \cite{ilvento2020implementing} showed how to implement the exponential mechanism while avoiding the pitfalls of finite precision arithmetic. Report noisy max with exponential noise can be implemented on a finite precision computer using the techniques of Canonne et al.~\cite{canonne2020discrete}.

Finally, we remark that lower bounds for the private selection problem are known \cite{steinke2017tight}. These show that the algorithms we have discussed are essentially optimal.

\section{Conclusions}\label{sec:conc}
In this paper, we showed that the recently proposed permute-and-flip mechanism is really report noisy max with exponential noise in disguise. This means that the optimality results obtained by McKenna and Sheldon \cite{paf} directly apply to the latter mechanism and that it should be adopted wherever feasible. 

There are some situations where the exponential mechanism is preferable. Such a situation is illustrated in work by Blocki et al.~\cite{BlockiDB16} in releasing a nondecreasing histogram, where the output space is too massive to enumerate. The simpler form  of the exponential mechanism output probabilities were important in the algorithm design. It is an open question of how to design corresponding algorithms so that they would be equivalent to the optimal version of report noisy max.


\bibliographystyle{plain}
\bibliography{refs}
\end{document}